\documentclass[aps,showpacs,superscriptaddress,nofootinbib,10pt]{revtex4}

\usepackage{graphicx}
\usepackage{dcolumn}
\usepackage{bm}

\begin{document}

\title{Effects of Instanton Induced Interaction on the Pentaquarks}

\author{Tetsuya Shinozaki}%
\email[e-mail: ]{shinozk@th.phys.titech.ac.jp}
\affiliation{%
Department of Physics, H-27, Tokyo Institute of Technology,\\ Meguro, Tokyo 152-8551, Japan
}%
\author{Makoto Oka}%
\affiliation{%
Department of Physics, H-27, Tokyo Institute of Technology,\\ Meguro, Tokyo 152-8551, Japan
}%
\author{Sachiko Takeuchi}%
\affiliation{%
Japan College of Social Work, Kiyose 204-8555, Japan
}%

\begin{abstract}
Roles of instanton induced interactions (III) in the masses of pentaquark baryons,  $\Theta^+$ ($J=1/2$ and 3/2) and $\Xi^{--}$, and a dibaryon, $H$, are discussed using the MIT bag model in the negative parity case. 
It is shown that the two-body terms in III give a strong attraction mainly due to the increase of the number of pairs in multi-quark systems. In contrast, the three-body $u$-$d$-$s$ interaction is repulsive.
It is found that III lowers the mass of $\Theta^+$ as much as 100 MeV from the mass predicted by the bag model without III.
\end{abstract}


\maketitle

\section{Introduction}

Reports of discoveries of exotic baryons, $\Theta^+$\cite{Nakano:2003qx} and $\Xi^{--}$\cite{Alt:2003vb}, started intensive discussions on various possibilities of bound pentaquark states.
Existence of the $\Theta^+$, in particular, has been confirmed by several groups\cite{exp}.
This is a baryon with strangeness $+1$ and therefore cannot be made of three quarks.
The minimal quark content of $\Theta^+$ is $u^2d^2\bar s$, and thus it is called 
``pentaquark''.

Properties of the new baryons have not yet been determined fully.  The most essential properties are spin and parity.  Predictions  by the chiral soliton model\cite{Diakonov:1997mm,Praszalowicz:2003ik} and various quark models\cite{Hosaka:2003jv,Stancu:2003if,Jennings:2003wz,Karliner:dt,Jaffe:2003sg,Glozman:2003sy,Kochelev:2004nd,Kanada-Enyo:2004bn,Hiyama04} claimed a $1/2^{+}$ state of $\Theta^+$, while constituent quark model with all the five quarks sitting in the lowest energy level predicts a negative parity ground state, $1/2^{-}$\cite{Carlson:2003pn}.  (The other possible spin is $3/2$, but most of the dynamical quark models, that can reproduce spectra of ordinary hadrons, predict that lower spin states have lower masses.)
Furthermore, majority of QCD-based calculations, such as QCD sum rule\cite{Sugiyama:2003zk} and lattice QCD\cite{Sasaki:2003gi,Mathur:2004jr,Ishii:2004qe}, indicate that the positive parity state has a higher mass.

There is, however, a strong argument that makes the positive parity state favorable.
The observed width of $\Theta^+$ decaying into $KN$ scattering states seems very small, i.e., likely to be less than 20 MeV.
It is hard to understand this small width if the parity of $\Theta^+$ is negative, because it then decays into a $KN$ S-wave state.  Recent calculation\cite{HOSAKA04} shows that $1/2^{-}$ state should have a width of order a few hundred MeV, while $1/2^{+}$ state may have much smaller width due mainly to the suppression by  the centrifugal barrier.

Several possibilities to reduce the mass of the positive parity state have been proposed in literatures\cite{Karliner:dt,Jaffe:2003sg,Teke04}. Among them, proposals of exotic clustering of a diquark or tri-quark inside the pentaquark system are popular. For instance, Jaffe and Wilczek\cite{Jaffe:2003sg} proposed a structure of the pentaquark as a bound state of two scalar $ud$ diquarks and $\bar s$. Because of the color structure and the Bose symmetry of diquarks, this bound state has to have a nonzero angular momentum, $L=1$, between the diquarks and thus the resulting $\Theta^+$ has positive parity.

Under these circumstances, it is important to study contributions of various different quark dynamics to the pentaquark states in detail. 
The constituent quark model has several important dynamical ingredients, among which we here consider confinement, perturbative one-gluon exchange interaction and nonperturbative instanton induced interaction. 

The first one has an obvious role, that is, to confine quarks inside a color singlet hadron.
In the nonrelativistic quark models, a confinement potential of a linear form is often used.
Its main role is to give localized wave functions of quarks in the hadron.
However, the confinement mechanism for pentaquarks has not been fully understood in potential quark models, 
while the confinement potential was calculated in lattice QCD\cite{Okiharu:2004wy} in the heavy quark limit.
The MIT bag model\cite{Chodos:1974je} gives another picture of confinement, that is, quarks are confined by the bag pressure at the bag surface.  The bag is a cavity of perturbative vacuum in the nonperturbative vacuum of QCD and thus has a larger volume energy.  This energy difference between the outside and the inside of the bag produces pressure at the bag surface and confines quarks.
We consider that this confinement picture is valid for the pentaquark systems because it reflects only the nature of the QCD vacua.  

Important roles of the one-gluon exchange (OgE) interactions for the meson and baryon spectra  were first pointed out by DeRujula, Goergi and Glashow\cite{DeRujula:1975ge}.  They showed in particular that the color-magnetic interaction is essential to explain the mass difference of the octet and decuplet baryons.  They, however, noticed that the spectrum of the pseudoscalar mesons, the large mass differences among $\pi$, $\eta$ and $\eta'$, is not reproduced purely by OgE. This difficulty is known to be related to the $U_A(1)$ anomaly, which was pointed out by Weinberg\cite{Weinberg:ui}.

The instanton induced interaction (III) may solve the $U_A(1)$ problem, as was shown by 't Hooft\cite{'tHooft:fv}.
The instanton is a solution of the Yang Mills equation in the Euclidean 4-dimensional space\cite{Belavin:fg}.
Its coupling to light quarks has special properties that it occurs through zero modes of the quark localized around the instanton and that it changes chirality and thus breaks $U_A(1)$ symmetry.
In the meson spectrum, III gives significant mass splitting between $\eta$ and $\eta'$.

Interesting roles of III in the baryon spectrum were studied by Shuryak-Rosner\cite{Shuryak:bf} and Oka-Takeuchi\cite{Oka:ud} in 1989.  They pointed out that the spin-spin interaction comes also from III and III (with or without OgE) gives a similar spectrum of baryons.  Therefore the baryon spectrum cannot be used to determine which of the two interactions are responsible for the mass splitting of hadrons.
It is, of course, most likely that both of them contribute to the octet-decuplet splitting. Indeed, if we assume that the splitting comes only from the color-magnetic interaction of OgE, then the coupling strength required to reproduce empirical data is in general too large in the sense that the perturbative treatment is hardly justified.  It is, therefore, favorable that III is responsible for a part of the octet-decuplet splitting and thus reduces the part given by OgE.

Other possibilities of inter-quark interactions include exchange of the Nambu-Goldstone bosons of chiral symmetry breaking\cite{Glozman:2003sy,Stancu:2003if,Teke04}. The NG boson exchange gives a flavor-spin dependent force, which contributes to the octet-decuplet mass splitting.  It was also pointed out that such interaction may generate considerable diquark correlations in the pentaquark baryon\cite{Teke04}. 
We, however, do not consider the NG boson exchange in the present study in the context of the MIT bag model.

The purpose of this paper is to clarify roles of III in the pentaquark systems employing the model mentioned above.
We consider III in the context of the MIT bag model. Namely, we assume that III is applied to quarks inside a bag.  This assumption may need some explanation.  In the naive bag model, the ``vacuum'' inside the bag is identified as the perturbative vacuum where there is no 
gluon condensate. As a consequence, instanton densities inside the bag could be different from that employed in the instanton vacuum picture.  It is, however, argued that such an approach is not able to reproduce appropriate spectrum of the pseudoscalar mesons. Therefore we consider a situation that the vacuum structure is minimally modified, while quantum excitation modes exist only in the bag.  On the other hand, it is natural to assume that the confinement mechanism that successfully describes three-quark baryon states is common to the pentaquark baryons. Also there is an advantage of the bag model, i.e., the quarks inside the baryon are treated relativistically.

We concentrate on the masses of the exotic baryons, and first point out that the mass of $\Theta^+$ predicted by quark models that reproduce the masses of ordinary (3-quark) baryons is considerably larger that the experimental value, 1540 MeV. This is demonstrated for the $1/2^-$  state in the bag model\cite{Carlson:2003pn}. Next, we advocate importance of III in the pentaquark systems\cite{Takeuchi:1993rs,shinozk04}. We stress that III becomes significant for flavor singlet states of quarks and exotic states with many quarks.

Some previous works consider III in the context of diquark models of the pentaquark baryons\cite{Kochelev:2004nd}. 
III is well known to give strong correlation between $ud$ quarks forming a scalar diquark with isospin $I=0$ and spin $S=0$.  It was also suggested that III is a main driving force of color superconductivity phase that may appear in QCD at a finite baryon density. Recent calculation shown important contribution from two-body correlation, although they include only the OgE interaction\cite{Kanada-Enyo:2004bn,Hiyama04,Teke04}. We, however, do not take into account two-body correlations in this work, but only single particle motions of the ``shell model'' type are considered, although inclusion of such correlation is the next step.

In sect. 2, we give formulations of the instanton induced interactions and the MIT bag model.
In sect. 3, the results for the masses and sizes of the pentaquarks and $H$ dibaryon are presented.
In sect. 4, the conclusion is given.

\section{Formalism}
\subsection{Instanton Induced Interaction}
The instanton induced interaction (III), introduced by 't Hooft\cite{'tHooft:fv}, is an interaction among quarks of $N_f\,(=3)$ light flavors,
which is induced by coupling of quarks to instantons (and anti-instantons) through zero modes.
The instanton is a classical solution of the gauge field equation for the Euclidean 4-dimensional space, which is one of the most important non-perturbative effects in QCD.
The main difference from the perturbative gluon-exchange interactions is that III is not chirally invariant and applies only on flavor singlet states of quarks. 
't Hooft pointed out that III solves the $U_A(1)$ problem, namely the origin of $\eta-\eta'$ mass splitting.
We employ a long wave length approximation so that the size of the instanton ($\simeq 1/3$ fm) is neglected.
Then the interaction is written as a contact interaction\cite{Oka:1990vx}.
\begin{eqnarray}
H^{(3)}&=&G^{(3)} \epsilon_{ijk} \epsilon_{i'j'k'} \, \bar{\psi}_{R,i}(1)\bar{\psi}_{R,j}(2)\bar{\psi}_{R,k}(3) \left( 1-\frac{1}{7} \sum_{i<j} \sigma_i \cdot \sigma_j \right)\, \psi_{L,k'}(3)\psi_{L,j'}(2)\psi_{L,i}(1) \nonumber\\
 &+& \mathrm{(h.c.)}, \label{H3}\\
H^{(2)}&=&G^{(2)} \epsilon_{ij} \epsilon_{i'j'}  \, \bar{\psi}_{R,i}(1)\bar{\psi}_{R,j}(2) (1-\frac{1}{5}\sigma_1 \cdot \sigma_2)\psi_{L,j'}(2)\psi_{L,i'}(1) + \mathrm{(h.c.)}, \label{H2}
\end{eqnarray}
where $\psi$ is the field operator of the quark, $\epsilon_{ijk}$ is totally asymmetric tensor, and $i$, $j$ and $k$ represent flavor.  $R$ and $L$ are chiral indices. $H^{(3)}$ is the 3-body Hamiltonian and $H^{(2)}$ is  the 2-body Hamiltonian obtained by contracting a quark pair in the 3-body III into a quark condensate or the quark mass term.
Then the relation between the strengths of the 2-body and 3-body III is given by
\begin{eqnarray}
G^{(2)}_{ud} \simeq \frac{25}{7} \frac{\langle \bar{u} u \rangle}{2}\frac{m_s^{\mathrm{eff}}}{m_u^{\mathrm{eff}}} G^{(3)},
\end{eqnarray}
where we neglect the current quark mass, $m^{\mathrm{eff}}$ is the constituent mass of the quark. We use $m_u^{\mathrm{eff}}/m_s^{\mathrm{eff}} \simeq 0.6$. The $\langle \bar{u} u \rangle$ is the quark condensate. We use  $\langle \bar{u} u \rangle \simeq (-225\mathrm{MeV})^3$. 
The 3-body interaction $H^{(3)}$ is  repulsive,  while the 2-body interaction $H^{(2)}$ is attractive  because the quark condensate is negative. The 2-body III has three flavor types, ($H^{(2)}_{ud},H^{(2)}_{us},H^{(2)}_{ds}$). $H^{(2)}_{ud}$, for instance, is an interaction between $u$ and $d$ quarks that is obtained by contracting $s\bar{s}$ quark pair from the 3-body Hamiltonian. As a result, $H^{(2)}_{ud}$ is proportional to the effective quark mass of strangeness, $m_s$. Thus we obtain
\begin{eqnarray}
G^{(2)}_{us} = \frac{m_u^{\mathrm{eff}}}{m_s^{\mathrm{eff}}}G^{(2)}_{ud}.
\end{eqnarray}

We concentrate on the ground state baryons.
Both the 2-body III and the 3-body III do not affect the decuplet baryons, $\Delta, \Sigma^{*}, \Xi^{*}$ and $\Omega$,
because III is applied only to flavor anti-symmetric quarks.
Likewise for the octet baryon, $N, \Lambda, \Sigma$ and $\Xi$, the 3-body III does not exert.
On the other hand, the 2-body III gives attractive force. 
But the effects of the  2-body III are similar to the one gluon exchange (OgE) regarding the spin structure\cite{Shuryak:bf}.
Thus the baryon spectrum can be reproduced by any combination OgE and the 2-body III, that is,
we can not determine the strength of III from the baryon spectrum.

Takeuchi and Oka\cite{Takeuchi:1990qj} pointed out that the 3-body III plays an important role in the $H$ dibaryon, a flavor singlet system of $u^2d^2s^2$. 
The fact that a deeply bound $H$ dibaryon does not exist favors a strong 3-body III, while is repulsive in the $H$ system.
The pseudoscalar and scalar mesons are also known to be sensitive to III\cite{Kunihiro:1987bb,Hatsuda:1994pi,Takizawa:1996nw,Naito:1999sb,Naito:2003zj}.

Since $\Theta^{+}$ has only one strange anti-quark, the repulsive force from the 3 body III is expected to be weaker than
that for the $H$ dibaryon.

\subsection{MIT Bag Model}
We introduce the instanton induced interaction in the MIT bag model. This effect for the baryon and meson is shown by Kochelev \textit{et al.}\cite{Kochelev:de}. The mass splitting between the octet baryons and the decuplet baryons comes from the hyperfine interaction. The origin of the hyperfine interaction can be either the one gluon exchange (OgE) or the instanton induced interaction (III).

The mass of a hadron in the MIT Bag Model\cite{DeGrand:cf} is given by 
\begin{eqnarray}
M(R)&=& n_u w(m_u,R) + n_s w(m_s,R)+\frac{4\pi}{3}B R^3 \nonumber\\
&-&\frac{Z_0}{R}+ (1 - P_{III}) \sum_{i>j} (\vec{\sigma_i} \cdot \vec{\sigma_j})\,(\vec{\lambda_i}\cdot \vec{\lambda_j}) M_{ij}(R)\nonumber\\
&+& P_{III}(H^{(3)}(R)+H^{(2)}(R)) + E_0 ,\label{bagmass}
\end{eqnarray}
where $R$ is the bag radius, $n_u$ is the number of the $u,d$ quarks, and $n_s$ is the number of the $s$ and $\bar{s}$ quarks. The function $w(m,R)$ denotes the single particle energy obtained as an eigenvalue of the Dirac equation  for the bag. 
$-Z_0/R$ is the zero point energy as well as corrections due to the center of mass motion of the bag. $B$ is the bag constant. The fifth term is the color-magnetic part from OgE. $M_{ij}(R)$ is the strength including the spatial contribution, which is given in Appendix \ref{ap:OgE}.
And we neglect the color-electric interaction.
The $H^{(3)}$ and $H^{(2)}$ are contributions of III which we introduce.
$P_{III}$ is a parameter which represents the portion of the hyperfine splitting induced by III.
By multiplying $P_{III}$ to III and $(1-P_{III})$ to OgE, we keep the mass splitting of $N-\Delta$ to 300MeV. 
If $P_{III}=0$, the mass splitting of $N-\Delta$ comes purely from OgE, while for $P_{III}=1$ it comes purely from III.
Another way of determining $P_{III}$ is to reproduce the $\eta-\eta'$ mass difference. 
Although the method suffers from model dependence, we estimated $P_{III}=0.25-0.6$, and 
in the case of the MIT Bag model, $P_{III}=0.31$\cite{Takeuchi:sg}.
$E_0$ in Eq.(\ref{bagmass}) is a parameter introduced to reproduce the mass of the nucleon.
If we ignore the changes of the radii of $N$ and $\Delta$ under the variation of $P_{III}$, $E_0$ is given roughly by $E_0 = 150 \mathrm{MeV} \times P_{III}$.
The effect of $E_0$ can be taken into account by changing $Z_0$ and $B$ accordingly, but here for simplicity we fix $Z_0$ and $B$ and change $E_0$ as a function of $P_{III}$.

We obtain the masses and the radii of the pentaquarks by minimizing $M(R)$,
\begin{eqnarray}
\frac{d M(R)}{d R} = 0.
\end{eqnarray}
The parameters of the bag model are taken from the original MIT bag model, DeGrand \textit{et al.}\cite{DeGrand:cf},
 $B = (0.145 \,\, \mathrm{GeV})^4$, $Z_0$ = 1.84, $\alpha_c$=$\alpha_s/4$ = 0.55, $m_s$ = 0.279 GeV and $m_u$ = 0 GeV.
It is found that the mass spectrum of the 3-quark baryons hardly change by introducing III.

We consider a negative parity pentaquark.
In the MIT bag model, such a state is given by placing all the quarks in the ground $S_{\frac{1}{2}}$ state,
which is obtained by solving the free Dirac equation with linear boundary condition $-i (\vec{\gamma} \cdot \vec{n}) \psi = \psi$ at $r=R$.
The single particle energy of this ground state is obtained by solving
\begin{eqnarray}
\tan{x} &=& \frac{x}{1- m R -\sqrt{x^2+m^2 R^2}},\\
w(m,R) &=& \sqrt{\left( \frac{x(m,R)}{R} \right)^2 + m^2}.
\end{eqnarray}
For the massless quark, $x$=2.04.
The wave function of the lowest energy $S_{\frac{1}{2}}$ state is given by 
\begin{eqnarray}
\psi  &=& \left(
\begin{array}{c}
 f(m, r) \\
 - i  \,(\vec{n} \cdot \vec{\sigma}) \,g(m, r)
\end{array}
\right)  \, \chi_{s} 
e^{- i  w_0 t},
\end{eqnarray}
where 
\begin{eqnarray}
f(m, r)&=&N
\sqrt{\frac{w_0 + m}{2 w_0}} j_0 \left( \frac{x}{R}r \right) ,\\
g(m, r)&=&-N
\sqrt{\frac{w_0 - m}{2 w_0}} j_1 \left( \frac{x}{R}r \right),
\end{eqnarray}
where $N$ denotes the normalization constant,  $R$ is the bag radius, $j_i$ is the spherical Bessel function, and $\chi_{s}$ denotes a two component Pauli spinor.
For the antiparticle, we get the negative frequency mode operated by charge conjugation,
\begin{eqnarray}
{\psi^c} &=&  C \bar{\psi}^t =  i \gamma^2 \gamma^0 {\bar{{\psi }}}^t.
\end{eqnarray}
We show the matrix elements of OgE and III in Appendix.

\section{Results}
We consider the pentaquarks  $\Theta^{+}$ composed of  $uudd\bar{s}$ with isospin 0, spin 1/2 and negative parity, 
and $\Xi^{--}$, a partner within the flavor $\bar{10}$ with isospin 3/2. 
We also consider $\Theta^{+}_{S=3/2}$, which is the spin partner of $\Theta^{+}$. Note that 
the only possible spin partner for the negative parity case is the spin 3/2 state, and  no spin 5/2 state is allowed due to permutation symmetry.

\begin{figure}
\begin{center}
\includegraphics[width=0.48\textwidth]{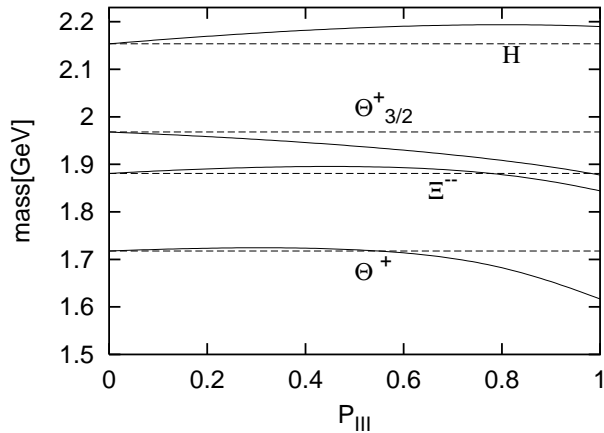}
\end{center}
\caption{\label{mass}The masses of $\Theta^{+}, \Xi^{--}, \Theta^{+}_{S=3/2}$, and $H$.
The dashed lines are the values at $P_{III}=0$.}
\end{figure}

In Fig. \ref{mass} and Table \ref{tab:mass}, we show the masses of the pentaquarks as functions of $P_{III}$.
The dashed lines are the values at $P_{III}=0$, which correspond to the masses under the influence only of OgE.
The right end, $P_{III}=1$, gives the masses when the $N-\Delta$ splitting is purely due to III. 
First, we point out that the pure OgE lowers the masses of $\Theta^{+}$ from the noninteracting 5 quark state.
One sees that the $\Theta^{+}$ is affected by III most strongly among these states.
At $P_{III}$=1, the mass of $\Theta^{+}$ is 100MeV smaller than that at $P_{III}=0$.
It is, however, not sensitive to $P_{III}$ in the region, $P_{III}=0.25 \sim 0.6$, which is considered to be realistic.
In contrast, $\Theta^{+}_{S=3/2}$ changes significantly in this region. 
For all $P_{III}$, the mass of $\Xi^{--}$ is almost constant. The mass of the $H$ dibaryon grows monotonically as $P_{III}$ increases.

It is found that the mass of $\Theta^{+}$ does not agree with the experimental value (1540MeV) even if the full III is introduced.
On the other hand, the model reproduces the mass of $\Xi^{--}$ at $P_{III}=0.25 \sim 0.6$.
In order to improve this situation we may further take into account an attractive interaction or correlation.

\begin{table}
\caption{\label{tab:mass}The masses and radii of the pentaquarks and the $H$ dibaryon.}
\begin{ruledtabular}
\begin{tabular}{ c c c c c }
                                        & $\Theta^{+}$	&$\Theta^{+}_{S=3/2}$&$\Xi^{--}$&  $H$     \\ \hline
$M(P_{III}=0)$[MeV]                     & 1717          &1968                &1881        & 2154 \\
$M(P_{III}=1)$[MeV]                     & 1617          & 1877               & 1844       & 2190 \\
$M(P_{III}=1)-M(P_{III}=0)$[MeV]        & $-101$        & $-91$              & $-37$      & $+36$ \\
$R(P_{III}=0)$[fm]                      & 1.14          & 1.22               & 1.13       & 1.20 \\
$R(P_{III}=1)$[fm]                      & 0.76          & 0.98               & 0.83       & 1.00 \\
\end{tabular}
\end{ruledtabular}
\end{table}

\begin{figure}
\begin{center}
\includegraphics[width=0.48\textwidth]{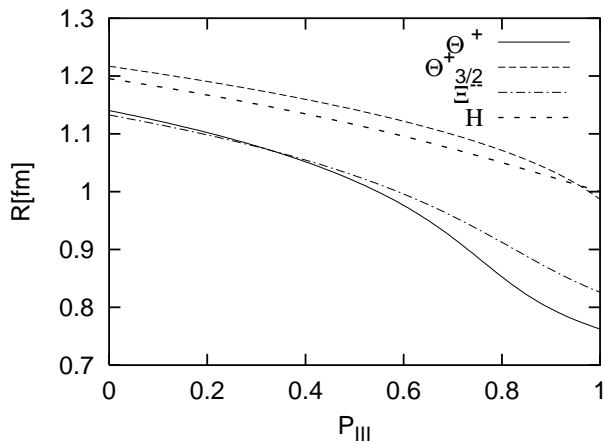}
\end{center}
\caption{\label{radius}The bag radii for the pentaquarks.}
\end{figure}

In Fig. \ref{radius} and the last two rows of Table \ref{tab:mass}, the radii of the considered baryons are given.
They show that $\Theta^{+}$  shrinks as $P_{III}$ increases. 
$\Theta^{+}$ and $\Xi^{--}$ have almost the same radii at $P_{III}=0$, which are larger than the radius of the nucleon 5 $\mathrm{GeV}^{-1}$. At the realistic region of $P_{III}$, 
the radii of pentaquarks are about $0\%\sim20\%$ larger than the radius of the nucleon.
We find that the strongly attractive force of III  does not reduce the masses, but rather shrinks the radii for the all states.

Contributions of each term of Eq.(\ref{bagmass}) are listed for at $P_{III}$=0, $\frac{1}{2}$ and 1 in Tables \ref{tab:piii0}, \ref{tab:piii05} and \ref{tab:piii1}.
The contribution of the 3 body III is roughly 10\% of that of the 2 body III for the pentaquarks.
For the $\Theta^{+}_{S=3/2}$, the contribution of OgE is very small.
Thus effects of III are most easily seen in $\Theta^{+}_{S=3/2}$. 
We find that the difference between $\Theta^{+}$ and $\Xi^{--}$ is roughly equal to the mass difference of the $u$ and $d$ quarks  and  the $s$ quark.
The 3 body interaction is stronger in the $H$ dibaryon than the other cases.
The main reason is that the $H$ dibaryon has more ($u d s$) combinations.

\begin{table}
\caption{\label{tab:piii0}Contributions of each term of the mass $M(R)$ at $P_{III}=0$ in units  of GeV.
B is for the term $\frac{4\pi}{3}BR^3$, $Z_0$ for $\frac{-Z_0}{R}$, and W for $n_u w_u + n_s w_s$.
Note that OgE, $H^{(2)}$ and $H^{(3)}$ correspond to the bare contributions, which are not multiplied by the $P_{III}$ factors,
$(1-P_{III})$ for OgE  and $P_{III}$ for III. The actual contributions are obtained by multiplying the $P_{III}$ factors.
}
\begin{ruledtabular}
\begin{tabular}{ c c c c c c c c c }
                    &  Total & B     &   $Z_0$  & W      &  OgE     &$H^{(2)}$ &$H^{(3)}$& $E_0$\\ \hline
$\Theta^{+}$        & 1.717  & 0.369 & $-0.318$ & 1.930  & $-0.264$ & $-0.397$ & 0.034   & 0    \\ 
$\Theta^{+}_{S=3/2}$& 1.968  & 0.449 & $-0.298$ & 1.821  & $-0.004$ & $-0.193$ & 0.007   & 0    \\ 
$\Xi^{--}$          & 1.881  & 0.362 & $-0.320$ & 2.108  & $-0.269$ & $-0.390$ & 0.038   & 0    \\ 
  $H$               & 2.154  & 0.425 & $-0.303$ & 2.354  & $-0.323$ & $-0.431$ & 0.054   & 0    \\
\end{tabular}
\end{ruledtabular}
\end{table}

\begin{table}
\caption{\label{tab:piii05}Contributions of each term of $M(R)$ at $P_{III}=0.5$ in units of GeV.}
\begin{ruledtabular}
\begin{tabular}{ c c c c c c c c c }
                    &  Total & B     &   $Z_0$  & W       &  OgE     &$H^{(2)}$&$H^{(3)}$& $E_0$\\ \hline
 $\Theta^{+}$       & 1.720 & 0.263 & $-0.356$  & 2.138   & $-0.301$ & $-0.554$ & 0.065  & 0.070 \\
$\Theta^{+}_{S=3/2}$& 1.938 & 0.371 & $-0.317$  & 1.928   & $-0.003$ & $-0.233$ & 0.019  & 0.070 \\
 $\Xi^{--}$         & 1.895 & 0.270 & $-0.353$  & 2.284   & $-0.301$ & $-0.518$ & 0.066  & 0.070 \\
  $H$               & 2.187 & 0.346 & $-0.325$  & 2.494   & $-0.350$ & $-0.527$ & 0.080  & 0.070 \\
\end{tabular}
\end{ruledtabular}
\end{table}

\begin{table}
\caption{\label{tab:piii1}Contributions of each term of $M(R)$ at $P_{III}=1$ in units of GeV.}
\begin{ruledtabular}
\begin{tabular}{ c c c c c c c c c}
                     &  Total & B     &   $Z_0$  & W       &  OgE     &$H^{(2)}$ &$H^{(3)}$& $E_0$\\ \hline
 $\Theta^{+}$        &1.617   & 0.110 & $-0.475$ & 2.795   & $-0.419$ & $-1.304$ & 0.351   & 0.140 \\
 $\Theta^{+}_{S=3/2}$&1.877   & 0.239 & $-0.367$ & 2.202   & 0.001    & $-0.358$ & 0.022   & 0.140 \\
 $\Xi^{--}$          &1.844   & 0.140 & $-0.439$ & 2.751   & $-0.386$ & $-0.984$ & 0.235   & 0.140 \\
 $H$                 &2.190   & 0.249 & $-0.362$ & 2.739   & $-0.398$ & $-0.727$ & 0.152   & 0.140 \\
\end{tabular}
\end{ruledtabular}
\end{table}

\begin{figure}
\begin{center}
\includegraphics[width=0.48\textwidth]{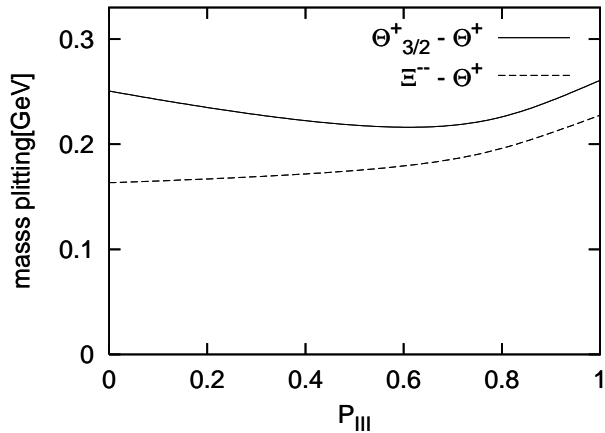}
\end{center}
\caption{\label{split}The mass splittings of $\Theta^{+}_{S=3/2} - \Theta^{+}$ and $\Xi^{--} - \Theta^{+}$.}
\end{figure}

Fig. \ref{split} shows the mass splitting between $\Theta^{+}$ and the other pentaquarks.
The mass splitting between $\Xi^{--}$ and $\Theta^{+}$ increases as $P_{III}$ because the single particle energy in the bag behaves as $\sim \frac{1}{R}$.
On the other hand, the mass splitting between $\Theta^{+}_{S=3/2}$ and $\Theta^{+}$ comes from the spin-spin interaction between the four quarks state with spin 1 and the $\bar{s}$ quark, and its dependence on $P_{III}$ is weak.

Finally, we show the contribution of each term of III. III in Eqs.(\ref{H3}) and (\ref{H2}) have two types of interactions, which are
the term proportional to identity and the term dependent on the spin. 
We show the numbers for the $\Theta^{+}$ with its bag radius same as the nucleon, R=5 $\mathrm{GeV}^{-1}$=0.985fm.
\begin{eqnarray}
\langle \phi_{\Theta^{+}}| H^{(2)} |\phi_{\Theta^{+}} \rangle &=& G^{(2)}_{ud} \, (18.6_1 + 1.9_\sigma) \times 10^{-3} ,\\
\langle \phi_{\Theta^{+}}| H^{(3)} |\phi_{\Theta^{+}} \rangle &=& G^{(3)} \, ( 12.2_1 - 3.2_\sigma ) \times 10^{-5}.
\end{eqnarray}
The total contribution, which is $H^{(2)}$ and $H^{(3)}$, is 
\begin{eqnarray}
\langle \phi_{\Theta^{+}}|H^{(2)} + H^{(3)}|\phi_{\Theta^{+}} \rangle &=& G_{ud}^{(2)} \, (14.5_1 + 2.8_\sigma)\times 10^{-3}. 
\end{eqnarray}
The numbers suffixed by 1 ($\sigma$) denote the contributions from the identity (spin) term of III.
We find that the main contribution of III in the pentaquark comes from the identity term, which gives $\simeq 83\%$ of the total value. 
The corresponding value for the nucleon is $5/8\simeq 63\%$. The contribution of the identity term is so large because its effect is essentially proportional to the number of quark pairs in the pentaquark, namely 10, while it is only 3 for the ordinary baryon.

\section{Conclusion}
Effects of the instanton induced interaction on the spectrum of $\Theta^+$, $\Theta^{+}_{S=3/2}$,  $\Xi^{--}$ and $H$ dibaryon have been studied using the MIT bag model of hadron structure in the negative parity case.
The bag provides a permanent confinement and the mass splittings are attributed to spin-dependent interactions induced by the one-gluon exchange as well as the instanton-quark couplings.

We have found that III lowers the mass of $\Theta^+$ and $\Theta^{+}_{S=3/2}$, while the mass of $H$ increases as the strength of III increases.  The $\Xi^{--}$ mass is insensitive to the strength of III. We have also found that the strongly attractive force of III makes the bag radii shrink. In fact, the radii of the pentaquarks happen to be as small as the radius of the 3-quark baryons. The attraction comes mainly from the spin-independent part of III, which is proportional to the number of quark (anti-quark) pairs in the system.  Thus the attraction in the pentaquark systems is (10/3) times as large as that in the three-quark baryons. The attraction is stronger for the $H$ dibaryon because the number of pairs is 15 in the six-quark system. It is, however, the 3-body III that strongly repulsive for $H$.  The repulsion is weaker in $\Theta^+$ (and in $\Theta^{+}_{S=3/2}$) than in $H$, because the number of $u$-$d$-$s$ combinations inside the pentaquark is just a half of that in $H$.

The 2-body III is weaker in the $\Xi^{--}$ system because the $H^{(2)}$ is proportional to the inverse of the effective quark masses and it is therefore weaker for the strange quark.  We have also found that the spin-spin splitting between the $\Theta^{+}$ and $\Theta^{+}_{S=3/2}$ is larger than the splitting of $\Theta^+$ and $\Xi^{--}$.  This is one of the distinctive feature of the negative parity pentaquarks in comparison to the positive parity states, where the spin-orbit force is responsible for the $J=1/2^{+}-3/2^{+}$ splitting and it is generally weaker than the spin-spin force.
Moreover, the contribution of OgE for the $\Theta^{+}_{S=3/2}$ is very small. Therefore the splitting of $\Theta^{+}$ and $\Theta^{+}_{S=3/2}$, if it is found, should give us important information on the structure of the pentaquarks.

In all, we have shown that the instanton induced interaction is an important ingredient of the pentaquark spectrum and should be taken into account in evaluating their masses. The present results are not completely satisfactory in detail, however. For instance, it can not reproduce the observed $\Theta^+$ mass.  It also underestimates the mass of $H$, which has not been found below the $\Lambda\Lambda$ threshold, 2231 MeV. Possible resolutions are corrections from expected two-body (diquark type) correlations, pionic effects, which may be included in chiral bag models, and also couplings to background $NK$ scattering states.  If (one or some of) these effects are important, the pentaquark spectrum may be well modified.  The instanton induced interaction will contribute to the two-body correlation significantly, as has been already suggested in relation to the color superconductivity of the QCD vacuum at finite baryon density.  It seems also important to include the effect of the $NK$ continuum, which should make the width of the negative parity pentaquark state large.

Despite these defects, the current study is worthwhile because using the simplest possible picture of the hadron, we demonstrate how large and important are the effects of instantons on the spectrum of pentaquarks and the $H$ dibaryon. Further analysis including the above-mentioned corrections are to be performed as the next step. 

\begin{acknowledgments}
This work is supported in part by the Grant for Scientific Research (B)No.15340072, (C)No.16540236 and (C)No.15540289 from the Ministry of Education, Culture, Sports, Science and Technology, Japan. T.~S. is supported by a 21st Century COE Program at Tokyo Tech "Nanometer-Scale Quantum Physics" by the Ministry of Education, Culture, Sports, Science and Technology.
\end{acknowledgments}

\appendix
\section{One Gluon Exchange}\label{ap:OgE}
We show the matrix elements of the color-magnetic part of the one-gluon exchange (OgE) between quarks. $M_{i j}$ denotes the strength (matrix element) for the quark pair of $i$ and $j$. For the ground state, the Hamiltonian is given by
\begin{eqnarray}
H_{OgE} &=& \sum_{i<j} (\vec{\lambda}_i \cdot \vec{\lambda}_j)\,(\vec{\sigma}_i \cdot \vec{\sigma}_j) M_{ij} ,
\end{eqnarray}
where 
\begin{eqnarray}
M_{ij} &=&  -3\,\alpha_c  \frac{\mu(m_i,R)\mu(m_j,R)}{R^3}I(m_i,m_j,R),
\end{eqnarray}
\begin{eqnarray}
I(m_i,m_j,R) &=& 1 + 2\int_0^R \frac{\mathrm{d}r}{r^4} \mu(m_i,r) \mu(m_j,r),\\
\mu(m, r) &=& \int_0^r \mathrm{d}r' \, \mu'(m , r'),\\
\mu'(m, r) &\equiv&  \frac{-2}{3}\,r^3\,f(m,r)\,g(m,r).
\end{eqnarray}
For anti-quark, we replace $\lambda$  by $-\lambda^{*}$.

For the octet baryons, OgE is attractive. For the decuplet baryons, it has the same strength but the sign is opposite,
\begin{eqnarray}
\langle \phi_{N}|H_{OgE} |\phi_{N} \rangle &=& 8 M_{uu},\\
\langle \phi_{\Delta}|H_{OgE}|\phi_{\Delta} \rangle &=&  -8  M_{uu}.
\end{eqnarray}
For the $H$ dibaryon, it is strongly attractive, as was first shown by Jaffe\cite{Jaffe:1976yi},
\begin{eqnarray}
\langle \phi_{H}| H_{OgE}|\phi_{H} \rangle &=& 5 M_{uu}  + 22 M_{us} -3 M_{ss}.
\end{eqnarray}
For the $\Theta^{+}$, we show the isospin decomposition,
\begin{eqnarray}
\langle \phi_{\Theta^{+}}|H_{OgE} |\phi_{\Theta^{+}} \rangle &=&  -\frac{26}{9}M_{uu}  - \frac{26}{9}M_{dd}  + \frac{100}{9} M_{ud} \nonumber\\
&+& \frac{20}{3}M_{us} +  \frac{20}{3}M_{ds}.
\end{eqnarray}
Then the matrix element of $\Theta^{+}$ is given by using $M_{uu}=M_{dd}=M_{ud}$ and $M_{us}=M_{ds}$. For the $\Xi^{--}$, we obtain a similar formula by $u \rightarrow s$ and $\bar{s} \rightarrow \bar{u}$:
\begin{eqnarray}
\langle \phi_{\Theta^{+}}|H_{OgE}  |\phi_{\Theta^{+}} \rangle &=&  \frac{16}{3} M_{uu} +  \frac{40}{3}M_{us}, \\
\langle \phi_{\Xi^{--}}| H_{OgE} |\phi_{\Xi^{--}} \rangle &=& \frac{34}{9}M_{uu} + \frac{160}{9}M_{us} - \frac{26}{9}M_{ss}.
\end{eqnarray}
For $\Theta^{+}_{S=3/2}$, we obtain
\begin{eqnarray}
\langle \phi_{\Theta^{+}_{S=\frac{3}{2}}}|H_{OgE}|\phi_{\Theta^{+}_{S=\frac{3}{2}}} \rangle &=& \frac{16}{3}M_{uu}- \frac{20}{3}  M_{us}.
\end{eqnarray}

\section{instanton induced interaction}
We show the matrix elements of the instanton induced interaction (III). For the octet baryon, III is attractive, while III does not affect the decuplet baryon:
\begin{eqnarray}
\langle \phi_{N}| H^{(2)} |\phi_{N} \rangle &=& \frac{G^{(2)}_{ud}}{4\pi} \int^{R}_{0} r^2 \mathrm{d}r \, \frac{24}{5}\left(f_u(r)^4 +  2f_u(r)^2 g_u(r)^2+g_u(r)^4 \right),\\
\langle \phi_{\Delta}| H^{(2)} |\phi_{\Delta} \rangle &=& 0.
\end{eqnarray}
For the $\Theta^{+}$,  we obtain
\begin{eqnarray}
\langle \phi_{\Theta^{+}}| {H}^{(2)}_{ud} |\phi_{\Theta^{+}} \rangle &=& \frac{G^{(2)}_{ud}}{4\pi}  \int^{R}_{0}  r^2 \mathrm{d}r \, \frac{4}{5}\left( 9 f_u(r)^4 +  2f_u(r)^2 g_u(r)^2 + 9g_u(r)^4 \right),\\
\langle \phi_{\Theta^{+}}|{H}^{(2)}_{us}|\phi_{\Theta^{+}} \rangle &=& \langle \phi_{\Theta^{+}}|{H}^{(2)}_{ds}|\phi_{\Theta^{+}} \rangle  = \frac{G^{(2)}_{us}}{4\pi} \int^{R}_{0}  r^2 \mathrm{d}r \,\frac{2}{15} \nonumber \\
& &\left(33f_u(r)^2 f_s(r)^2+33g_u(r)^2 g_s(r)^2-13f_u(r)^2 g_s(r)^2 \right.\nonumber\\ 
&-&\left. 13g_u(r)^2 f_s(r)^2  + 52f_u(r)g_u(r)f_s(r)g_s(r) \right),\\
&&\langle \phi_{\Theta^{+}}| {H}^{(3)} |\phi_{\Theta^{+}} \rangle = \frac{G^{(3)}}{(4\pi)^2} \int^{R}_{0}  r^2 \mathrm{d}r \,   \nonumber\\
&& \frac{60}{7}\left(
3f_u(r)^4 f_s(r)^2 - f_u(r)^4 g_s(r)^2 + 8f_u(r)^3 g_u(r) f_s(r) g_s(r)  \right. \nonumber\\
&-& 4f_u(r)^2 g_u(r)^2 f_s(r)^2 +4f_u(r)^2 g_u(r)^2  g_s(r)^2 \nonumber \\
&-&\left. 8f_u(r) g_u(r)^3 f_s(r) g_s(r) + g_u(r)^4 f_s(r)^2 - 3g_u(r)^4 g_s(r)^2
\right).
\end{eqnarray}

For $\Theta^{+}_{S=3/2}$, the matrix element of $H_{ud}^{(2)}$ is identical to that of $\Theta^{+}$:
\begin{eqnarray}
\langle \phi_{\Theta^{+}_{S=3/2}}| {H}^{(2)}_{ud} |\phi_{\Theta^{+}_{S=3/2}} \rangle &=& \frac{G^{(2)}_{ud}}{4\pi}  \int^{R}_{0}  r^2 \mathrm{d}r \, 
 \frac{4}{5}\left( 9 f_u(r)^4+ 2 f_u(r)^2 g_u(r)^2 + 9 g_u(r)^4 \right),\\
\langle \phi_{\Theta^{+}_{S=3/2}}|{H}^{(2)}_{us}|\phi_{\Theta^{+}_{S=3/2}} \rangle &=& \langle \phi_{\Theta^{+}_{S=3/2}}|{H}^{(2)}_{ds}|\phi_{\Theta^{+}_{S=3/2}} \rangle  =  \frac{G^{(2)}_{us}}{4\pi} \int^{R}_{0}  r^2 \mathrm{d}r \nonumber\\
&& \left( \frac{7}{5}f_u(r)^2 f_s(r)^2+\frac{7}{5}g_u(r)^2 g_s(r)^2-\frac{41}{15}f_u(r)^2 g_s(r)^2 \right. \nonumber\\
&-& \left. \frac{41}{15} g_u(r)^2 f_s(r)^2  + \frac{4}{3}f_u(r)g_u(r)f_s(r)g_s(r) \right),\\
\langle \phi_{\Theta^{+}_{S=3/2}}| {H}^{(3)} |\phi_{\Theta^{+}_{S=3/2}} \rangle &=& \frac{G^{(3)}}{(4\pi)^2}\int^{R}_{0} r^2 \mathrm{d}r \nonumber\\
&& \frac{15}{7}\left( 
3f_u(r)^4 f_s(r)^2 - 7 f_u(r)^4 g_s(r)^2 + 8f_u(r)^3 g_u(r) f_s(r) g_s(r)  \right. \nonumber\\
&+&  2 f_u(r)^2 g_u(r)^2 f_s(r)^2 -2 f_u(r)^2 g_u(r)^2  g_s(r)^2 \nonumber\\
&-&\left.8f_u(r) g_u(r)^3 f_s(r) g_s(r) + 7 g_u(r)^4 f_s(r)^2 - 3 g_u(r)^4 g_s(r)^2
\right).
\end{eqnarray}
For $\Xi^{--}$, we obtain
\begin{eqnarray}
\langle \phi_{\Xi^{--}}| {H}^{(2)}_{ud} |\phi_{\Xi^{--}} \rangle &=&  \frac{G^{(2)}_{ud}}{4\pi} \int^{R}_{0}  r^2 \mathrm{d}r 
 \left( \frac{22}{5}f_u(r)^4+\frac{22}{5}g_u(r)^4 + \frac{52}{15}f_u(r)^2 g_u(r)^2  \right) , \\
\langle \phi_{\Xi^{--}}| ({H}^{(2)}_{us+ds}) |\phi_{\Xi^{--}} \rangle &=& \frac{G^{(2)}_{us}}{4\pi} \int^{R}_{0}  r^2 \mathrm{d}r \nonumber\\
&&\left( \frac{58}{5}f_u(r)^2 f_s(r)^2+\frac{58}{5}g_u(r)^2 g_s(r)^2-\frac{202}{45}f_u(r)^2 g_s(r)^2 \right.\nonumber\\
&-&\left. \frac{202}{45} g_u(r)^2 f_s(r)^2  + \frac{632}{45}f_u(r)g_u(r)f_s(r)g_s(r) \right),\\
\langle \phi_{\Xi^{--}}| {H}^{(3)} |\phi_{\Xi^{--}} \rangle  &=& \frac{G^{(3)}}{(4\pi)^2} \int^{R}_{0}  r^2 \mathrm{d}r \nonumber\\ 
&&\frac{20}{21}\left( 27 {f_u(r)}^4 {f_s(r)}^2 - 14 {f_u(r)}^4 {g_s(r)}^2 + 28 { f_u(r)}^3 g_u(r) f_s(r)  g_s(r) \right.\nonumber\\
&+&13{ f_u(r)}^2 { g_u(r)}^2 { f_s(r)}^2   -13{ f_u(r)}^2 { g_u(r)}^2 { g_s(r)}^2 \nonumber\\
&-&\left.28f_u(r) { g_u(r)}^3 f_s(r)  g_s(r)  + 14{ g_u(r)}^4 { f_s(r)}^2 -27{ g_u(r)}^4 { g_s(r)}^2  \right) .
\end{eqnarray}
For the $H$ dibaryon, we obtain
\begin{eqnarray}
\langle \phi_{H}| {H}^{(2)}_{ud} |\phi_{H} \rangle &=& \frac{G^{(2)}_{ud}}{4\pi} \int^{R}_{0} r^2 \mathrm{d}r \, \frac{4}{5}(9 f_u(r)^4 + 2 f_u(r)^2 g_u(r)^2+ 9 g_u(r)^4),\\
\langle \phi_{H}| {H}^{(2)}_{us} |\phi_{H} \rangle &=& \langle \phi_{H}| {H}^{(2)}_{ds} |\phi_{H} \rangle = \frac{G^{(2)}_{us}}{4\pi}\int^{R}_{0} r^2 \mathrm{d}r \nonumber\\
&& \frac{2}{5}\left( 18f_u(r)^2 f_s(r)^2+18g_u(r)^2 g_s(r)^2-7f_u(r)^2 g_s(r)^2 \right.\nonumber\\
&-&\left. 7g_u(r)^2 f_s(r)^2  + 18f_u(r)g_u(r)f_s(r)g_s(r) \right) ,\\ 
\langle \phi_{H}| {H}^{(3)} |\phi_{H} \rangle &=& \frac{G^{(3)}}{(4\pi)^2} \int^{R}_{0} r^2 \mathrm{d}r \nonumber\\
&&\frac{30}{7}\left( 12f_u(r)^4 f_s(r)^2  - 12g_u(r)^4 g_s(r)^2 - 3 f_u(r)^4 g_s(r)^2 \right. \nonumber\\
 &+& 3 g_u(r)^4f_s(r)^2 + f_u(r)^2 g_u(r)^2 f_s(r)^2 - f_u(r)^2 g_u(r)^2 g_s(r)^2  \nonumber\\
&+& \left.14 f_u(r)^3 g_u(r) f_s(r) g_s(r)  - 14  f_u(r) g_u(r)^3 f_s(r) g_s(r) \right).
\end{eqnarray}

\end{document}